\documentclass[aps,pra,twocolumn,showpacs,superscriptaddress,floatfix,nofootinbib]{revtex4-1}

\usepackage{graphicx}
\usepackage{amssymb}
\usepackage{natbib}
\usepackage{amsmath}
\usepackage{etoolbox}
\usepackage{color}

\begin{document}

\title{Modeling magneto-optical trapping of CaF molecules}

\author{M. R. Tarbutt}
\email{m.tarbutt@imperial.ac.uk}
\affiliation{Centre for Cold Matter, Blackett Laboratory, Imperial College London,
Prince Consort Road, London SW7 2AZ, United Kingdom}
\author{T. C. Steimle}
\affiliation{Department of Chemistry and Biochemistry,
Arizona State University
Tempe, AZ 85287-1604}

\begin{abstract}
Magneto-optical trapping forces for molecules are far weaker than for alkali atoms because the photon scattering rate is reduced when there are multiple ground states, and because of optical pumping into dark states. The force is further reduced when the upper state has a much smaller Zeeman splitting than the lower state. We use a rate model to estimate the strength of the trapping and damping forces in a magneto-optical trap (MOT) of CaF molecules, using either the A$^{2}\Pi_{1/2}$ - X$^{2}\Sigma^{+}$ transition or the B$^{2}\Sigma^{+}$ - X$^{2}\Sigma^{+}$ transition. We identify a new mechanism of magneto-optical trapping that arises when, in each beam of the MOT, two laser components with opposite polarizations and different detunings address the same transition. This mechanism produces a strong trapping force even when the upper state has little or no Zeeman splitting. It is the main mechanism responsible for the trapping force when the A$^{2}\Pi_{1/2}$ - X$^{2}\Sigma^{+}$ transition is used.

\end{abstract}
\date{\today}
\maketitle

\section{Introduction}

There is currently great interest in cooling molecules to very low temperatures, motivated by a diverse range of applications \cite{Carr(1)09}. Laser cooling, which has been used to cool atoms to ultracold temperatures for decades, is difficult to apply to molecules because it is necessary to address multiple vibrational branches, each requiring a separate laser. Despite this difficulty, laser cooling has been demonstrated for the diatomic radicals SrF \cite{Shuman(1)10, Barry(1)12}, YO \cite{Hummon(1)13} and CaF \cite{Zhelyazkova(1)14}, and most recently a magneto-optical trap (MOT) of SrF molecules was demonstrated \cite{Barry(1)14, McCarron(1)15}. In all these cases, the laser cooling transition was the A$^{2}\Pi_{1/2}$ - X$^{2}\Sigma^{+}$ transition. The mechanism of magneto-optical trapping for these diatomic molecules has been elucidated in \cite{Tarbutt(1)15} where rate equations are used to model the interaction of the multi-level molecules with multiple frequencies of laser light.

For a two-level atom of mass $M$, wavelength $\lambda$ and decay rate $\Gamma$, the maximum achievable acceleration is $h \Gamma/(2M\lambda)$. Using the parameters of the A$^{2}\Pi_{1/2}$ - X$^{2}\Sigma^{+}$ transition of CaF, this is 290,000\,m\,s$^{-2}$. For the real molecule the acceleration is greatly reduced. This is partly because the molecule has multiple levels in the ground state that all need to be driven with separate laser frequencies, and this reduces the maximum achievable scattering rate relative to that of a two-level system \cite{Tarbutt(1)13}. This is compounded by optical pumping into states that are dark to the polarization of the laser beam that pushes displaced molecules back to the centre. Finally, even in the absence of these other difficulties, the trapping forces are very weak when the upper state $g$-factor, $g_{u}$, is much smaller than the lower state $g$-factor, $g_{l}$ \cite{Tarbutt(1)15}.
In the Hund's case (a) limit a $^{2}\Pi_{1/2}$ state has a negligibly small $g$-factor and so is unsuitable for making a MOT. The observed MOT of SrF was attributed to the mixing of the A$^{2}\Pi_{1/2}$ state with the B$^{2}\Sigma^{+}$ state, resulting in a value of $g_{u}$ that is large enough for weak trapping \cite{Tarbutt(1)15}. Magneto-optical trapping of CaF is being pursued, but given the very small value of $g_{u}$ in the A$^{2}\Pi_{1/2}$ state \cite{Devlin(1)15} it is unclear whether the A$^{2}\Pi_{1/2}$ - X$^{2}\Sigma^{+}$ transition is suitable for the MOT, or whether it is better to use the B$^{2}\Sigma^{+}$ - X$^{2}\Sigma^{+}$ transition. In this paper we explore this question and we identify a new mechanism of magneto-optical trapping that can give strong trapping forces irrespective of the upper-state $g$-factor, and that also side-steps the problem of optical pumping into dark states. We focus on CaF, though our conclusions also apply to other molecules of interest.

\section{Model}

We model magneto-optical trapping of CaF using a rate equation approach identical to that of \cite{Tarbutt(1)15}, except that here we use the full non-linear Zeeman shifts discussed below, rather than assuming linear shifts. We also use the same notation as in \cite{Tarbutt(1)15}. The angular momenta of the lower and upper states are written as $F_{l}$ and $F_{u}$ respectively, their projections onto the $z$-axis are $M_{l}$ and $M_{u}$, and in the linear Zeeman shift approximation the $g$-factors are $g_{l}$ and $g_{u}$. Along each coordinate axis of the MOT, molecules interact with counter-propagating beams, one pushing towards the centre (the restoring beam) and the other away from the centre (the anti-restoring beam). In a local coordinate system, which is always aligned to the direction of the magnetic field at the position of the molecule, the polarization of the restoring beam is $\sigma \pm$ according to whether it excites $\Delta M_{F} = \pm 1$ transitions. Throughout the discussion we take the field gradient in the $z$-direction, which is along the axis of the magnetic field coils, to be 20\,G/cm. The intensity distribution of the laser beams is Gaussian with a $1/e^{2}$ radius of 12\,mm.

A useful way to summarize the trapping force of the MOT is to calculate the acceleration of a stationary molecule versus its displacement along the $z$-axis, and the acceleration of a molecule at the centre of the MOT versus its speed in the $z$-direction. We call these acceleration curves. For small values of the displacement, $z$, and speed, $v_{z}$, we can write the acceleration as $a_{z}=-\omega_{z}^{2} z - \beta v_{z}$, where $\omega_{z}/(2\pi)$ is the trap frequency and $\beta$ is the damping coefficient. Their values provide a simple and convenient way of characterizing the MOT. To find the capture velocity of the MOT, $v_{c}$, we consider molecules entering the MOT in the xy-plane and at 45$^{\circ}$ to the laser beams, and calculate the fastest speed a molecule can have if it is to be captured. We consider MOTs using both the (0-0)A$^{2}\Pi_{1/2}$ - X$^{2}\Sigma^{+}$ and (0-0)B$^{2}\Sigma^{+}$ - X$^{2}\Sigma^{+}$ transitions, which we call the `A-state MOT' and `B-state MOT' respectively. The lower level is the first rotationally excited state of X$^{2}\Sigma^{+}$ (i.e. $N=1$), while the upper level is the lowest energy state of positive parity in either the A$^{2}\Pi_{1/2}$ or B$^{2}\Sigma^{+}$ states. Electric dipole selection rules for angular momentum and parity ensure that these states can only decay back to the $N=1$ rotational state of X$^{2}\Sigma^{+}$, meaning that these transitions are `rotationally closed'. None of the A$^{2}\Pi_{3/2}$ - X$^{2}\Sigma^{+}$  transitions have this property, which is why we do not consider this case.

\section{Hyperfine structure and Zeeman splitting}

\begin{figure}[t]
 \centering
 \includegraphics[width=0.45\textwidth]{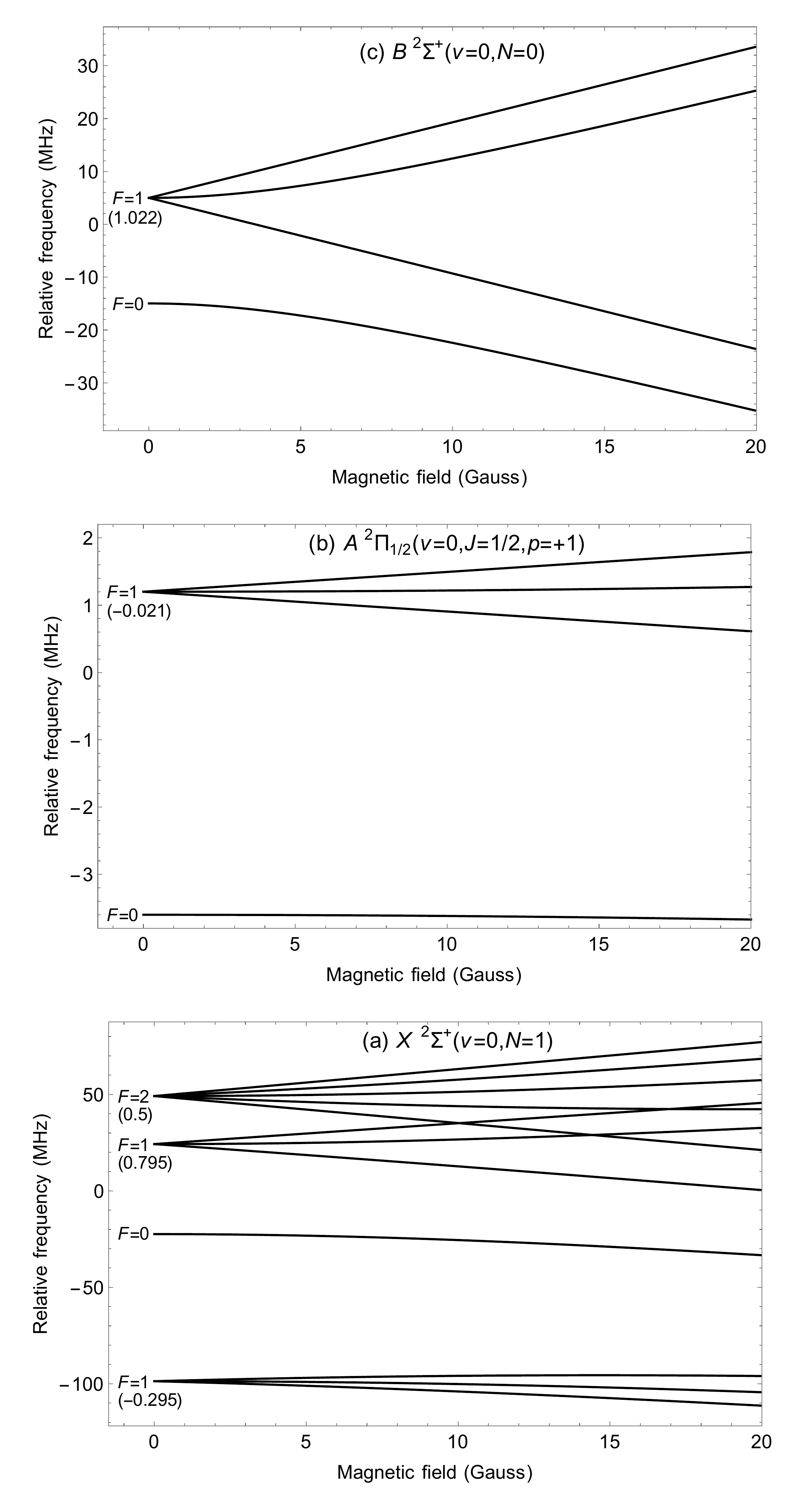}
\caption{\label{zeeman} Zeeman tuning of the states involved in magneto-optical trapping of CaF. (a) The four spin-rotation and hyperfine components of the X$^{2}\Sigma^{+} (v=0, N=1)$ state. (b) The two hyperfine components of the A$^{2}\Pi_{1/2}(v=0, J=1/2)$ state of positive parity ($e$-parity). (c) The two hyperfine components of the B$^{2}\Sigma^{+}(v=0, N=1)$ state. The numbers in brackets are the $g_{F}$-factors for each level.
}
\end{figure}

As we will see, the best scheme for making a MOT depends sensitively on the hyperfine structure and Zeeman splitting of the ground and excited states. Figure \ref{zeeman} shows the relevant energy levels of the X$^{2}\Sigma^{+}$, A$^{2}\Pi_{1/2}$ and B$^{2}\Sigma^{+}$ states, and their Zeeman tuning for magnetic fields up to 20\,G, calculated using the Zeeman parameters measured in \cite{Devlin(1)15}. At sufficiently low fields, the Zeeman shifts are linear, $\Delta E = g_{F}\mu_{B} M_{F} B$, and the values of $g_{F}$ are given in Fig.\,\ref{zeeman}. In the X$^{2}\Sigma^{+}(N=1)$ state, spin-rotation and hyperfine interactions lead to the four components shown in Fig.\,\ref{zeeman}(a), with total angular momenta $F=$1, 0, 1, and 2.  In the MOT, all four components need to be addressed. The field-free intervals between these levels are calculated using the spin-rotation and hyperfine parameters given in \cite{Childs(1)81}. The Zeeman tuning is significantly non-linear for fields above a few Gauss, and there are level-crossings above 10 Gauss due to the small energy interval between the upper $F=1$ and $F=2$ levels.

Figure \ref{zeeman}(b) shows the much smaller Zeeman tuning of the A$^{2}\Pi_{1/2} (J=1/2)$ state of positive parity (i.e. the $e$-parity component). The hyperfine splitting of this state is unknown. We have attempted to measure it by laser spectroscopy using the apparatus described in \cite{Devlin(1)15}. The hyperfine splitting of the $e$-parity and $f$-parity components of the A$^{2}\Pi_{1/2}(J=1/2)$ level are, in terms of the Frosch and Foley parameters, $2/3(h_{1/2}-d)$ and $2/3(h_{1/2}+d)$, respectively, where $h_{1/2} = a-(b+c)/2$ and $b=b_{F} - c/3$. Both $h_{1/2}$ and $d$ should be small and positive ($\sim$10\,MHz) given the nature of the A$^{2}\Pi$ state. Scans over the $P_{1}+Q_{12}(1)$, $Q_{1}(0)$, and $P_{12}(2)$ lines of the $(0,0)A^{2}\Pi_{1/2}- X^{2}\Sigma^{+}$ band, which probe the $e$-parity and $f$-parity components of the A$^{2}\Pi_{1/2}(J=1/2)$ level,  were recorded at a resolution of 20\,MHz FWHM and revealed no evidence of excited state hyperfine splitting. Similarly, the $^{S}R_{21}(0)$ and $P_{2}(3)$ lines of the $(0,0)A^{2}\Pi_{3/2} - X^{2}\Sigma^{+}$ band, which probe the $e$- and $f$-parity components of the $A^{2}\Pi_{3/2}(J=3/2)$ level, were recorded and show no evidence of excited state hyperfine structure. The observed and predicted spectra, and associated energy level diagrams, are given in the Supplemental Material. These measurements demonstrate that $h_{1/2}$, $h_{3/2}(=a+(b+c)/2)$ and $d$ are all less than 20\,MHz, and that the hyperfine interval in the A$^{2}\Pi_{1/2} (J=1/2)$ state is less than 10\,MHz. In our model, we set this interval to 4.8 MHz, which is the value determined for the hyperfine splitting of the corresponding negative parity level \cite{Wall(1)08}.

Figure \ref{zeeman}(c) shows the Zeeman tuning of the B$^{2}\Sigma^{+}(N=0)$ state. The hyperfine splitting of this state was also previously unknown, and so we have attempted to measure it. Scans over the $^{R}Q_{12}(1)$ and $P_{1} + ^{P}Q_{11}(1)$ lines of the $(0,0)B^{2}\Sigma^{+} - X^{2}\Sigma^{+}$ band were recorded at a resolution of 20\,MHz FWHM, and they reveal some small splittings and shifts due to excited state hyperfine structure. The observed and predicted spectra, and associated energy level diagram, are given in the Supplemental Material. From this data we estimate $b_{F}=20\pm5$\,MHz and $c=50\pm 10$\,MHz. This gives a 20\,MHz hyperfine interval for the B$^{2}\Sigma^{+}(N=0)$ state, which is the value we use in our model. Because of the small hyperfine splitting and rapid Zeeman tuning, the electron and nuclear spin decouple at low magnetic fields, and then the states are best characterized by the $M_{J}$ and $M_{I}$ quantum numbers.

\section{Dual-frequency MOT}
\label{Sec:Dual}

\begin{figure*}[!t]
 \centering
 \includegraphics[width=0.95\textwidth]{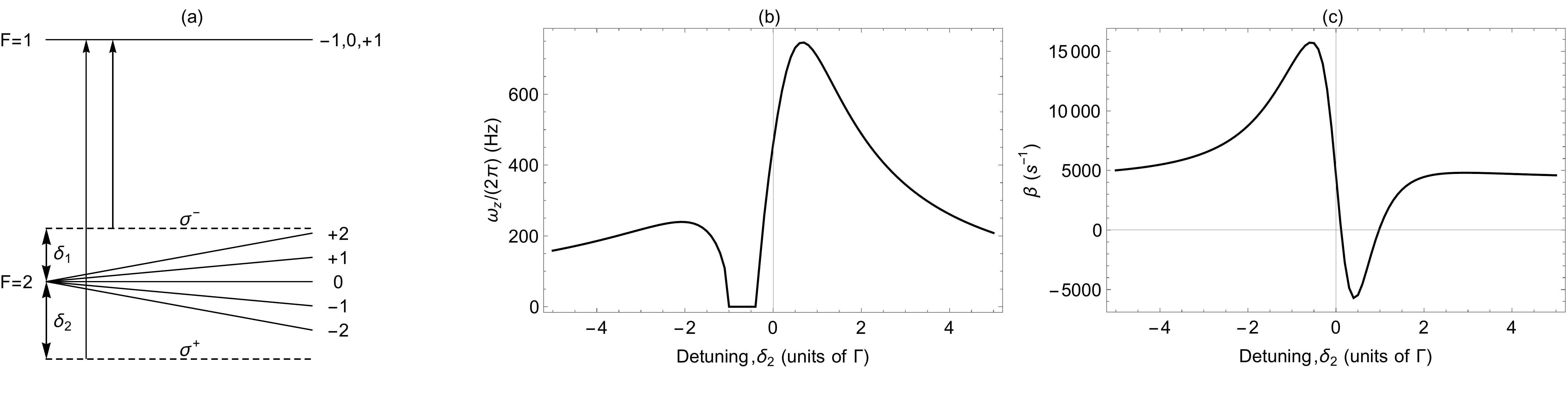}
\caption{\label{2f} (a) Illustration of a model MOT where the lower level has $F_{l}=2$, $g_{l}=0.5$, and the upper level has $F_{u}=1$, $g_{u}=0$.The transition is driven by two oppositely polarized frequency components with detunings of $\delta_{1}$ and $\delta_{2}$, indicated by the dashed lines. (b) Trapping frequency versus $\delta_{2}$. (c) Damping coefficient versus $\delta_{2}$.
}
\end{figure*}

We identify a new mechanism of magneto-optical trapping which avoids the problem of weak trapping associated with dark states and with a small $g_{u}$, and which is relevant to the rest of our discussion. Consider the simple system illustrated in Fig.\,\ref{2f}(a), which consists of a single lower level with $F_l=2$, $g_l=0.5$, an upper level with $F_u=1$, $g_u=0$, and two oppositely polarized frequency components with detunings $\delta_{1}$ and $\delta_{2}$. The wavelength, linewidth and mass are set equal to our CaF system. The molecule interacts with counter-propagating laser beams along each of the coordinate axes, each beam containing these two frequency components. The size of the laser beams is the same as described above, and the power per beam and per component is set to 40\,mW. We fix $\delta_{1} = -\Gamma$, and vary $\delta_{2}$. Figure \ref{2f}(b) shows the trapping frequency as a function of $\delta_{2}$. When $\delta_{2}$ is positive, meaning that one component is red-detuned and the other is blue-detuned, there is a restoring force. When $M_F$ is positive the Zeeman shift brings the transition into resonance with the red-detuned light, and so we have arranged for the red-detuned component of the restoring beam to drive $\Delta M_F=-1$ transitions so that the molecule preferentially interacts with that restoring beam (the red-detuned component of the anti-restoring beam drives $\Delta M_F=+1$ transitions). When $M_{F}$ is negative the transition comes into resonance with the blue-detuned frequency component, and because it has the opposite polarization the molecule again preferentially scatters from the restoring beam. The restoring force is maximized when $\delta_{2}$ is about $0.7\Gamma$, but remains significant even for large positive detunings. There is no trapping when $\delta_2$ is between $-0.3\Gamma$ and $-\Gamma$ because then the transitions are preferentially driven by the anti-restoring beams. However, for more negative detunings there is once again a restoring force. In this case, the restoring beam has one component driving $\Delta M_F=-1$ transitions, and another driving $\Delta M_F=+1$ transitions, with the former closer to resonance. For the anti-restoring beam, it is the opposite. When $M_F$ is positive, only $\Delta M_F=-1$ transitions are driven, and this is far more likely to be driven by the restoring beam since it is (relatively) much closer to resonance. When $M_F$ is negative, only $\Delta M_F=+1$ transitions are driven, and this is only a little more likely to be driven by the anti-restoring beam since the relative difference in detuning is not so great.

Figure \ref{2f}(c) shows how the damping coefficient depends on $\delta_2$. Cooling occurs when $\beta$ is positive. When $\delta_2$ is negative there is always cooling because both frequency components are red-detuned. When $\delta_2$ is positive and smaller than $\Gamma$ there is heating because the blue-detuned component is closer to resonance than the red-detuned one. When $\delta_2 > \Gamma$ there is cooling again, because the red-detuned component is the one closer to resonance. Together, these plots show that there is both cooling and trapping for all values of $\delta_2$ apart from values between 0 and $-\Gamma$. It also demonstrates the complexity that can arise even in this relatively simple system.

\section{The A$^{2}\Pi_{1/2}$ - X$^{2}\Sigma^{+}$ transition}

\begin{figure*}[!t]
 \centering
 \includegraphics[width=0.95\textwidth]{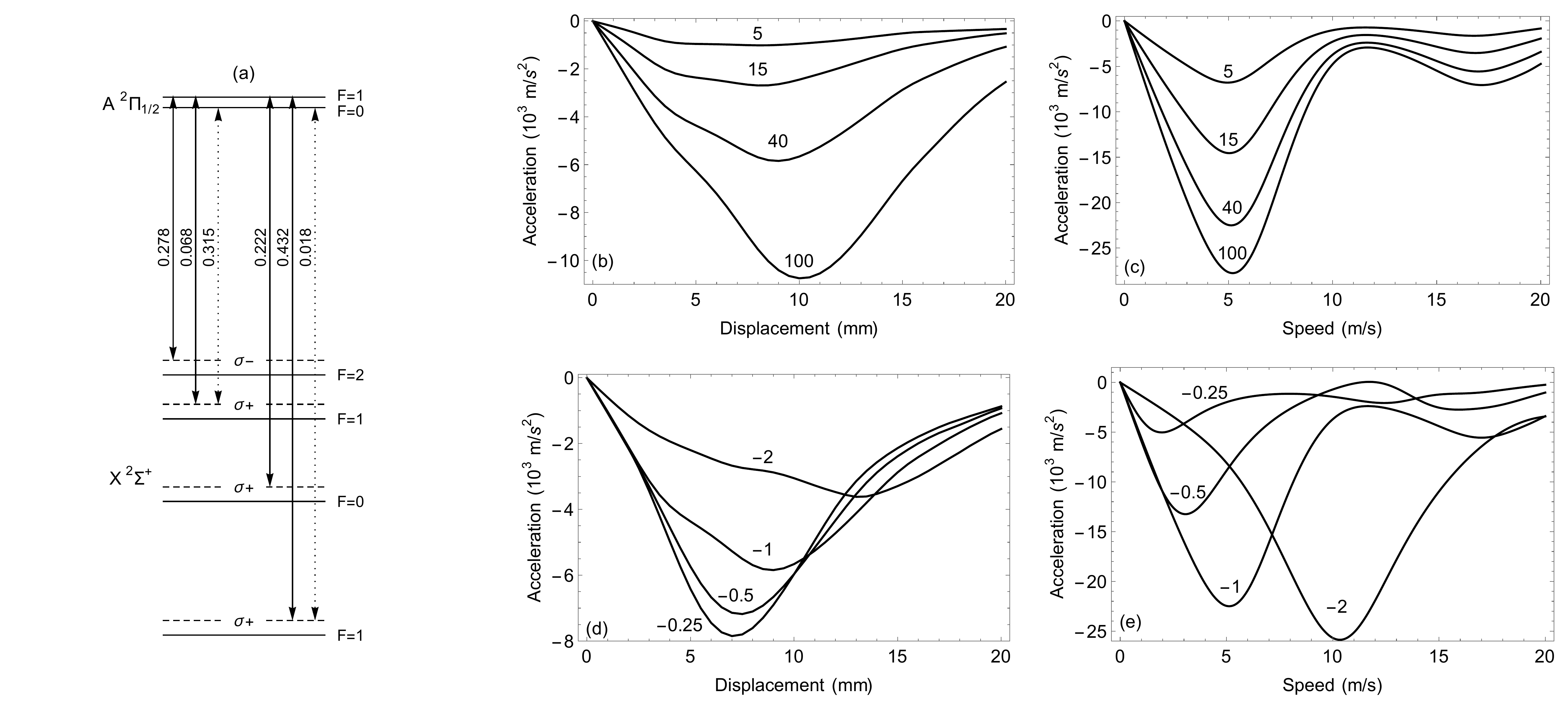}
\caption{\label{accelA}  CaF MOT operating on the A$^{2}\Pi_{1/2}$  - X$^{2}\Sigma^{+}$ transition. (a) The four frequency components of the laser have a common detuning, $\delta$, from $F_{l}$ to $F_{u}=1$, as shown by the solid arrows. Dotted arrows show allowed transitions to $F_{u}=0$. Values to the left of the arrows give the relative transition strengths for a fixed polarization, summed over $M_{u}$ and $M_{l}$. These are independent of the choice of polarization. The polarizations chosen for each component of the restoring beam are indicated as $\sigma\pm$. Acceleration versus (b) displacement and (c) speed, for four different values of the power in each MOT beam and each frequency component: 5, 15, 40 and 100 mW. The detuning is $-\Gamma$. Acceleration versus (d) displacement and (e) speed, for four different values of the detuning: $-0.25\Gamma$, $-0.5\Gamma$, $-\Gamma$ and $-2\Gamma$. The power is 40 mW.
}
\end{figure*}

We turn now to a model of a CaF MOT using the (0,0)A$^{2}\Pi_{1/2}$ - X$^{2}\Sigma^{+}$ transition - the A-state MOT. This transition has $\lambda = 606$\,nm and $\Gamma = 2\pi\times 8.3$\,MHz \cite{Wall(1)08}. Four frequency components are used to drive the four hyperfine components of the transition, and their polarizations in the restoring beam are chosen as shown in Fig.\,\ref{accelA}(a), which follows the recommendations of \cite{Tarbutt(1)15} for our case where $g_{u}$ is negative. We neglect the small branches to higher lying vibrational states.

Figure \ref{accelA}(b) shows how the acceleration of a stationary molecule depends on position for various values of the power per beam and per frequency component. Here we have set the detuning of all components to $-\Gamma$. We expect the MOT forces to be weak because of the small value of $g_{u}$, but we see from the figure that this is not the case. This is because the MOT forces are complicated by the close spacing of the upper $F_{l}=1$ and $F_{l}=2$ levels of the X$^{2}\Sigma^{+}$ state. This frequency interval is 25\,MHz, which is only $3\Gamma$. With our detuning of $-\Gamma$, the field-free transition from $F_l=2$ is driven (mainly) by two oppositely polarized components, one detuned by $-\Gamma$ and the other by $2\Gamma$. The polarizations of these components are exactly as in the model system of Fig.\,\ref{2f}(a), and Fig.\,\ref{2f}(b) tells us that this configuration produces a trapping force. Near the centre of the MOT it is this transition which is mainly responsible for the restoring force. Further from the centre, there is a strong additional trapping force coming from the upper $F_l=1$ transition. At a field of about 30\,G the $M_F=+1$ component of this transition comes into resonance with the laser component that is meant to drive $F_l=2$, while the $M_F=-1$ component comes into resonance with the laser component intended for $F_l=0$. Once again, the two laser components are polarized correctly so that the transitions from $M_F=+1$ and $M_F=-1$ are both driven by the restoring beam. This contribution to the trapping force pushes the turning point of the acceleration curve further out from the centre of the MOT, generating a large MOT capture volume. We also see from Figure \ref{accelA}(b) that the acceleration continues to increase as the laser power is increased up to 100\,mW.

Figure \ref{accelA}(c) shows how the acceleration of a molecule at the MOT centre depends on its speed, for the same set of laser powers as before. We see that this damping force peaks when the speed is near 5\,m/s, corresponding to a Doppler shift that equals the detuning of $-\Gamma$. There is also a secondary peak where the Doppler shift is $-4\Gamma$, which is again the result of there being two frequency components spaced by $3\Gamma$. This acts to broaden out the acceleration curve so that the damping force remains significant over a larger range of speeds.

Figures \ref{accelA}(d) and (e) show how the acceleration depends on position and speed for various values of the detuning, with the power fixed at 40\,mW. Bringing the detuning close to zero maximizes the trapping force in this case, but a negative detuning is required for there to be a damping force and so there is a trade-off between cooling and trapping. When the detuning is $-\Gamma$ and the power is 40\,mW we find $\omega_z=2\pi\times 160$\,Hz, $\beta = 5600$\,s$^{-1}$ and $v_c=20$\,m/s.

We attribute the strong restoring forces shown in Fig.\,\ref{accelA}(b,d) almost entirely to the dual-frequency effect. To verify this, we make the intervals between the four components of the transition much larger than $\Gamma$, setting the detuning to $-\Gamma$ and the power per beam and per frequency to 40\,mW. Then, we find a very simple dependence of the MOT forces on $g_{u}$: over the range $0\le g_u \le 0.14$, the maximum acceleration is $|a_{z,max}|=(17000 {\rm m/s}^2)|g_u|$ and the squared trapping frequency is $\omega_z^2 = (2\pi\times 282\,{\rm Hz})^2 |g_u|$. Note that the sign of $g_u$ dictates only the required sign of the field gradient. These relationships could also be applied to other similar molecules, suitably scaled by the ratio of the maximum scattering rates. For our present case, $g_u=-0.021$, and so the maximum acceleration is only 357\,m/s$^{2}$ and the trap frequency is 41\,Hz. We find similar results if we set the intervals between the four components equal to their true values, and also set the polarization handedness of the laser components all equal instead of the way they are shown in Fig.\,\ref{accelA}(a). These results demonstrate the large increase in the MOT forces brought about by the dual-frequency arrangement.

\begin{figure}[t]
 \centering
 \includegraphics[width=0.4\textwidth]{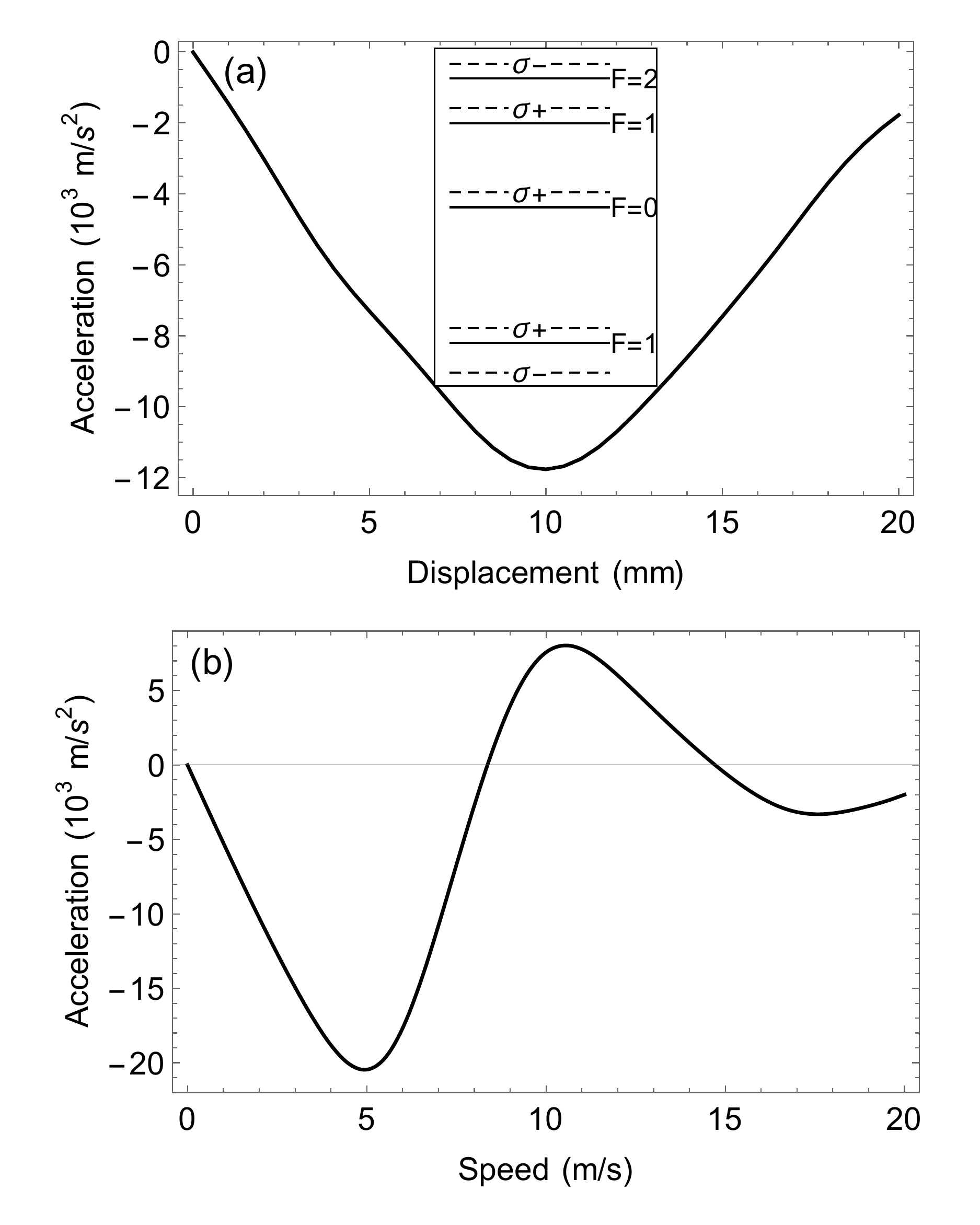}
\caption{\label{accelAExtra} Acceleration versus (a) displacement and (b) speed, for a CaF MOT operating on the A$^{2}\Pi_{1/2}$ - X$^{2}\Sigma^{+}$ transition using the set of detunings and polarizations illustrated in the inset to (a) and discussed in the text. The power per beam and per frequency is 40\,mW, and the components are detuned by $-\Gamma$ apart from the additional component whose detuning is $2\Gamma$.
}
\end{figure}

It is interesting to consider whether the trapping forces can be increased even further by applying the dual frequency method to several of the hyperfine components. Figure \ref{2f} shows that when $\delta_{1}=-\Gamma$ and $\delta_{2}=2\Gamma$, the trapping and damping forces are both strong, so we can aim to arrange this situation for several of the hyperfine components. Because the $F_l=2$ and upper $F_l=1$ components are spaced by $3\Gamma$ we automatically have this ideal situation for the $F_l=2$ state, as discussed above, but this means that we cannot {\it also} have it for the upper $F_l=1$ state. The $F_l=0$ state has no Zeeman splitting and no dark states, so there is nothing to be gained from applying two oppositely polarized frequency components. This leaves the lower $F_l=1$ state. Because this has a negative $g$-factor, the Zeeman shift of the $M_{F}$ states are opposite to the way they are drawn in Fig.\,\ref{2f}(a), and so the polarizations of the two frequency components need to be reversed relative to that figure. These arguments lead us to add one more frequency to address the lower $F_l=1$ level, detuned by 2$\Gamma$ from this level and polarized $\sigma-$, giving us the set of five frequencies and polarizations illustrated in the inset of Fig.\,\ref{accelAExtra}. Figure \ref{accelAExtra}(a) gives the acceleration versus position when the laser power is 40\,mW in each beam and frequency component. The addition of the extra component more than doubles the maximum acceleration to 12000\,m\,s$^{-2}$, and the trap frequency increases to $\omega_{z}=2\pi\times 191$\,Hz. Figure \ref{accelAExtra}(b) shows the acceleration versus speed for this case. We find that the additional component has little effect on the damping coefficient or the maximum damping force, but it does narrow down significantly the range of velocities where there is damping. In fact, the force has the wrong sign for speeds between 8 and 15\,m/s because, when in the lower $F_l=1$ level, the molecule is Doppler shifted into resonance with the extra frequency component of the {\it co-propagating} beam. As a result, the capture velocity decreases to $v_c=14$\,m/s. These results suggest that a good strategy may be to first load the MOT using four frequency components, and then switch on the fifth to provide tighter confinement.

\section{The B$^{2}\Sigma^{+}$ - X$^{2}\Sigma^{+}$  transition}

\begin{figure*}[t]
 \centering
 \includegraphics[width=0.95\textwidth]{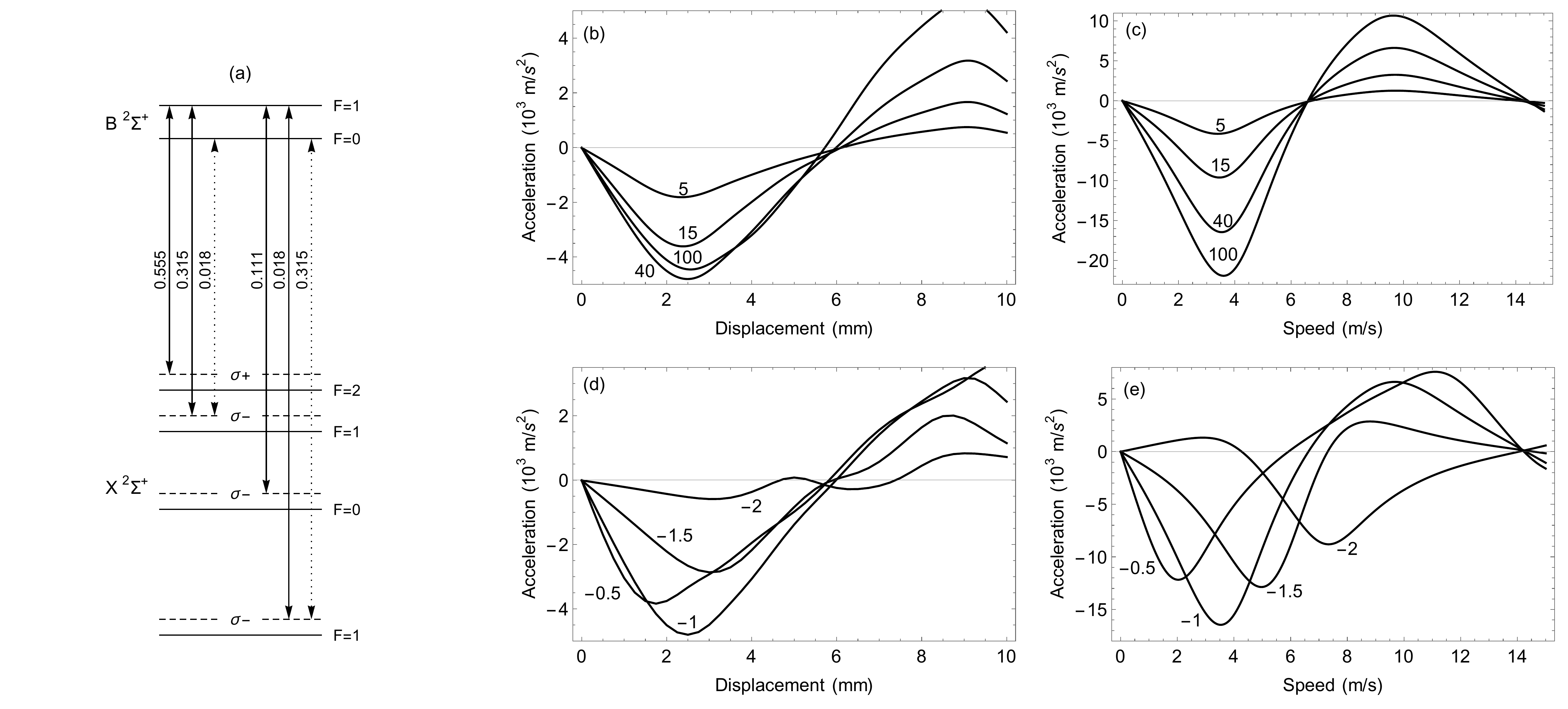}
\caption{\label{accelB} CaF MOT operating on the B$^{2}\Sigma^{+}$  - X$^{2}\Sigma^{+}$ transition. (a) The four frequency components of the laser have a common detuning, $\delta$, from $F_{l}$ to $F_{u}=1$, as shown by the solid arrows. Dotted arrows show allowed transitions to $F_{u}=0$. Values to the left of the arrows give the relative transition strengths for a fixed polarization, summed over $M_{u}$ and $M_{l}$. These are independent of the laser polarization. The polarizations chosen for each component of the restoring beam are indicated as $\sigma\pm$. (b) Acceleration versus displacement and (c) acceleration versus speed, for four different values of the power in each MOT beam and each frequency component: 5, 15, 40 and 100 mW. The detuning is $-\Gamma$. (d) Acceleration versus displacement and (e) acceleration versus speed, for four different values of the detuning: $-0.5\Gamma$, $-\Gamma$, $-1.5\Gamma$ and $-2\Gamma$. The power is 40 mW.
}
\end{figure*}

Next, we model a MOT using the (0-0)B$^{2}\Sigma^{+}$ - X$^{2}\Sigma^{+}$ transition - the B-state MOT. This transition has $\lambda=531$\,nm and $\Gamma = 2\pi \times 6.4$\,MHz. Population that leaks into other vibrational states of X$^{2}\Sigma^{+}$ can be rapidly pumped back into the cooling cycle via the A$^{2}\Pi_{1/2}$ state. These re-pumping transitions contribute very little to the cooling and trapping forces of the MOT, and so we neglect them in our model. Four laser frequency components address the four transitions from the X$^{2}\Sigma^{+}$ state, all with a common detuning. Their polarizations in the restoring beam are indicated in Fig.\,\ref{accelB}(a), following the recommendations of \cite{Tarbutt(1)15} for a positive $g_{u}$. We can choose whether to excite the transition to $F_{u}=0$ or $F_{u}=1$, which are spaced by 20\,MHz. The transition strengths are given in Fig.\,\ref{accelB}(a). We note that $F_{u}=0$ decays to the lower $F_{l}=1$ level 95\% of the time. If the laser is tuned close to resonance with this transition, the combination of the resonance condition and the stronger transition strength strongly favors excitation to $F_{u}=0$, and so these two levels can form a nearly-closed transition. Our model shows that this produces a high scattering rate and strong damping, but contributes little to the trapping force. Therefore, we choose to drive all $F_{l}$ components to $F_{u}=1$.

Figure \ref{accelB}(b) shows the acceleration versus position of stationary molecules for four different values of the power in each beam and frequency component. The detuning is $-\Gamma$. The curves have turning points near the position where the largest Zeeman shifts are equal to this detuning. Further out, the acceleration crosses zero because the Zeeman shifts become comparable to the upper state hyperfine interval and because of level crossings between the $F_{l}=1$ and 2 levels. The peak acceleration roughly triples as the power is increased from 5 to 40 mW, but does not increase much for powers exceeding 40 mW. This differs from the A-state MOT which requires more power to obtain the highest force. For the same set of powers, Fig.\,\ref{accelB}(c) shows the acceleration versus speed for a molecule at the centre of the MOT. The curves are similar to those shown in Fig.\,\ref{accelA}(c) for the A-state MOT, except that the range of speeds that give a damping force is reduced. The light is red-detuned by 6.4\,MHz for excitation to $F_{u}=1$, and so it is blue-detuned by 13.6\,MHz for excitation to $F_{u}=0$. When the speed is 7\,m/s the Doppler shift is 13\,MHz, bringing the {\em co-propagating} light into resonance with the transitions to $F_{u}=0$, and thus making the net force positive at this speed. This will limit the capture velocity of the MOT.

Figure \ref{accelB}(d) shows acceleration versus position for various detunings, $\delta$, with the power per beam and frequency component set to 40 mW. As $|\delta|$ increases the turning point of the acceleration curve shifts to larger displacements, because this turning point occurs roughly where the Zeeman shift matches the detuning. The peak restoring force is greatest when the detuning is approximately $-\Gamma$. A smaller value of $|\delta|$ reduces this peak force, just as it does in a simple one dimensional theory of the MOT \cite{Metcalf}. A larger value of $|\delta|$ also reduces the peak restoring force, which differs from the simple MOT theory, probably because of the different Zeeman tunings of the states involved. Figure \ref{accelB}(e) shows the acceleration versus speed for the same set of detunings. We note that a larger detuning shifts the point where the acceleration crosses zero to higher speeds. When the detuning is $-\Gamma$ and the power is 40\,mW we find $\omega_{z}=2\pi\times 254$\,Hz, $\beta = 4600$\,s$^{-1}$ and $v_c=10$\,m/s. Increasing the detuning to $-1.5\Gamma$ gives a smaller trap frequency and damping constant, $\omega_{z}=2\pi\times 166$\,Hz, $\beta = 1600$\,s$^{-1}$, but the capture velocity increases slightly, to 11\,m/s, due to the increased range of speeds where there is damping.

\begin{figure}[tb]
 \centering
 \includegraphics[width=0.4\textwidth]{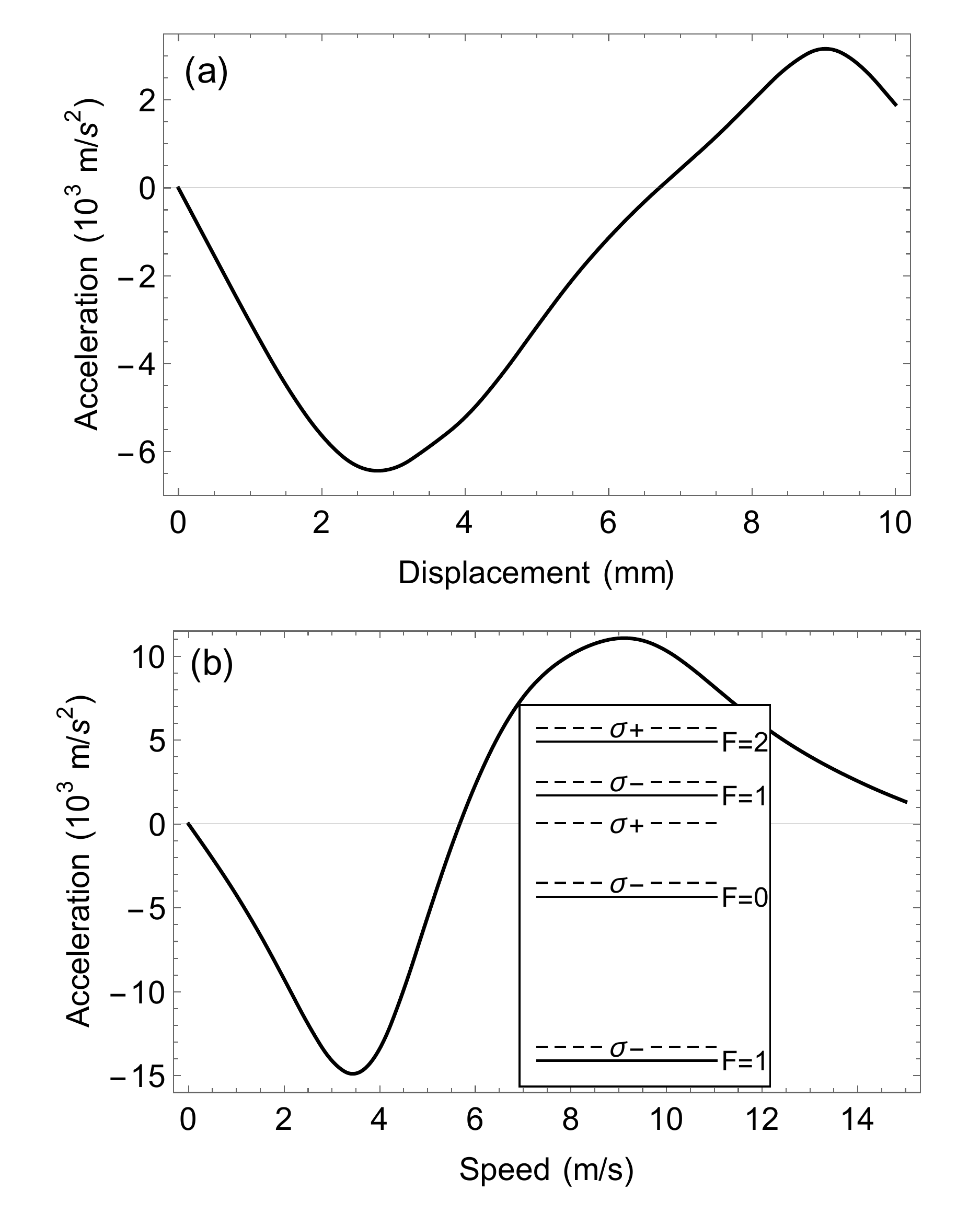}
\caption{\label{accelBExtra} Acceleration versus (a) displacement and (b) speed, for a CaF MOT operating on the B$^{2}\Sigma^{+}$ - X$^{2}\Sigma^{+}$ transition using the set of detunings and polarizations illustrated in the inset to (b) and discussed in the text. The power per beam and per frequency is 40\,mW, and the components are detuned by $-\Gamma$ apart from the additional component whose detuning is $2\Gamma$.
}
\end{figure}

When we increase the X$^{2}\Sigma^{+}$ hyperfine splittings by a factor of 10 from their true values, keeping the laser detunings the same as before, the acceleration curves do not change much. This shows that the dual-frequency effect discussed in Sec.\,\ref{Sec:Dual} does not play a major role in this MOT. As for the A-state MOT, it is worth trying to engineer a stronger trapping force by using the dual-frequency effect. Our modelling suggests that a good way to do this is to add one extra frequency component with $\sigma+$ polarization, detuned by $2\Gamma$ from the upper $F_{l}=1$ level. Figure \ref{accelBExtra} shows the acceleration versus position and speed for this arrangement. The detunings and polarizations are illustrated in the inset. With 40\,mW in each frequency component, the addition of this extra component increases the maximum trapping force by about 50\%, and the trap frequency increases to $\omega_{z}=2\pi\times 278$\,Hz. The damping coefficient and maximum damping force are almost unchanged. However, the damping force now has the wrong sign for all speeds between 5.7 and 15\,m/s, and this positive force is larger than before. This is due to the Doppler shift bringing the extra frequency component of the co-propagating beam into resonance, in addition to the the transitions to $F_{u}=0$ being Doppler-shifted into resonance as explained above. Consequently, the capture velocity decreases to $v_c=8$\,m/s.

\section{Summary}
Let us now summarize our findings. We find similar trapping forces in the A-state MOT and the B-state MOT. For the B-state MOT, when the detuning is $-\Gamma$  and the power is 40\,mW, the maximum acceleration of a stationary molecule is 4800\,m/s$^{2}$, the trapping frequency is $\omega_z = 2\pi\times 254$\,Hz, the damping coefficient is $\beta=4600$\,s$^{-1}$, and the capture velocity is $v_c=10$\,m/s. For the A-state MOT, with the same parameters, the maximum acceleration of a stationary molecule is 5800\,m/s$^{2}$, $\omega_z=2\pi\times 160$\,Hz, $\beta=5600$\,s$^{-1}$ and  $v_c=20$\,m/s. For the A-state MOT the forces act over a wider range of displacements and velocities, which is why it has a higher capture velocity. Throughout, we have assumed that vibrational re-pump transitions do not share the same upper state as the main MOT transition, in which case their influence on the MOT forces is small. When a re-pump transition does share the same upper-state as the MOT transition, we find that the force is reduced to about 70\% of the values found here.

We have identified a mechanism of magneto-optical trapping that arises when two laser components with opposite polarizations and different detunings address the same transition. The forces produced are strong even when the upper state has little or no Zeeman shift, and also when there are dark states amongst the lower levels. In the B-state MOT it is the large value of $g_u$ that contributes most to the trapping force, with the dual-frequency effects playing only a minor role. By contrast, the trapping force in the A-state MOT comes almost entirely from these dual-frequency effects. These effects also play a large role in the SrF MOT that has recently been demonstrated \cite{Barry(1)14, McCarron(1)15}. If we turn these effects off in our model of the SrF MOT (by artificially increasing the hyperfine intervals) the maximum acceleration is reduced by a factor of six, the turning point of the acceleration curve is a factor of two closer to the centre, and the trapping frequency is reduced by a factor of two. The dual-frequency effect can be utilized to produce stronger trapping forces using a carefully designed set of laser frequencies. For CaF, adding a single extra frequency component increases the restoring force by 100\% for the A-state MOT and by 50\% for the B-state MOT. However, the extra frequency has the effect of reducing the range of velocities where the damping force is strong, and this reduces the MOT capture velocity. The extra frequency component could be added to tighten the confinement once the MOT has already been loaded.

\acknowledgments This work has been supported by the UK Engineering and
Physical Sciences Research Council, under grants EP/I012044/1 and EP/M027716/1. The work at ASU was supported by the National Science Foundation, Division of Chemistry, CHE-1265885.

\end{document}